# Spatial heterogeneity in the radiogenic activity of the lunar interior: inferences from CHACE and LLRI on Chandrayaan-1

R. Sridharan[1]*, Tirtha Pratim Das, S.M.Ahmed[2], Gogulapati Supriya, Anil Bhardwaj

Space Physics Laboratory, VSSC, Indian Space Research Organisation, Trivandrum, India, PIN 695 022

and

J.A. Kamalakar

Laboratory for Electro Optical Systems, Bangalore, Indian Space Research Organisation, India

- [1] currently CSIR-ES at Physical Research Laboratory, Ahmedabad, India
- [2] currently at Central University Hyderabad, India
- * corresponding author: email: r_sridharan777@yahoo.co.in, sridharan@prl.res.in

Tele 91 79 26314777, fax: 91 79 26314659

## Abstract

In the past, clues on the potential radiogenic activity of the lunar interior have been obtained from the isotopic composition of noble gases like Argon. Excess Argon (40) relative to Argon (36), as compared to the solar wind composition, is generally ascribed to the radiogenic activity of the lunar interior. Almost all the previous estimates were based on, 'on the spot' measurements from the landing sites. **Relative concentration of the isotopes of $^{40}$Ar and $^{36}$Ar along a meridian by the Chandra's Altitudinal Composition Explorer (CHACE) experiment, on the Moon Impact Probe (MIP) of India's first mission to Moon, has independently yielded clues on the possible spatial heterogeneity in the radiogenic activity of the lunar interior in addition to providing indicative 'antiquity' of the lunar surface along the ground track over the near side of the moon**. These results are shown to **broadly** corroborate the independent topography measurements by the Lunar Laser Ranging Instrument (LLRI) in the main orbiter Chandrayaan-1. The unique combination of these experiments provided high spatial resolution data **while indicating** the **possible** close linkages between the lunar interior and the lunar ambience.

**Introduction**

The isotopes of Argon viz., $^{40}$Ar and $^{36}$Ar are considered to be very important constituents of the lunar atmosphere. Because of their inert nature and of being least affected by the out gassing from the space craft subsystems, they are considered to belong to sources outside the space craft including the lunar ambience. Almost in all past missions, wherever there had been some doubt about out-gassing from the vehicle/lander-module contaminating the in situ measurements, only those results pertaining to the noble gases have been reported in the literature (e.g., Hoffman etal 1973, Hodges and Hoffman 1975, Hodges 1973, Hodges 1975, Hodges 1978). Out of these two prominent isotopes of Argon, $^{40}$Ar is believed to be generated due to the radioactive decay of $^{40}$K inherently present in the rocks while $^{36}$Ar is considered to be produced by nucleosynthesis of stellar material, in this case, from the Sun. The situation is more analogous to planet Earth wherein $^{40}$Ar is one of the significant constituents of the Earth's atmosphere. Till recently, the most reliable measurements from the moon were from the LACE (Lunar Atmospheric Composition Experiment) of Apollo-17 as a part of which, a magnetic deflection type mass spectrometer was placed on the lunar surface by the astronauts. Unfortunately, LACE ran into saturation as the instrument drifted from the anti-sunward side to the sunward side during the course of one lunation. Though reliable composition measurements were made by LACE at least during times when the instrument was in the anti sunward side for several lunations, most of these results were kept aside due to scepticism that they could still be due to contaminants carried along with in addition to those from the landing module and other associated subsystems left on the lunar surface. The noble gases like He, Ne, Ar, Xe etc., were considered exceptions as they could not have been carried along, even by the remotest chances. The summary of the important findings have been succinctly provided by Stern (1999) and Wieler and Heber (2003). The conclusion that $^{40}$Ar is of lunar origin is essentially based on the relative concentration of $^{40}$Ar / $^{36}$Ar. This ratio has to be <<1 if these two isotopes are of solar wind origin. On the other hand, LACE measurements from the surface of the moon revealed that, the ratio remained >>1 all the times, and in fact, typically varied from 7.5-10 over the Apollo landing sites.

Once significant quantity of $^{40}$Ar has been measured along with the identification of a plausible source region within the lunar interior, several models were formulated (Hodges etal 1974, Hodges and Hoffman 1975, Hodges 1975). These models accounted for cold trap condensation/adsorption of Argon on the lunar surface along with one of the permanent loss processes viz., ionization due to solar wind energetic particles and the resultant Argon ions being swept away by the solar wind itself. Attempts were also made to take into account of Argon ions produced by the solar UV which then were accelerated in the electromagnetic field of the solar wind towards the lunar surface eventually getting implanted in the Lunar regolith (Manka and Michel 1971). However, all the earlier measurements were from fixed locations. Basically the earlier modelling attempts were aimed at explaining the diurnal variation exhibited by Argon during the Apollo mission. Most of the results corresponded to the first nine lunations during 1973, beyond which the LACE stopped functioning. Further, as mentioned earlier, these data are restricted only to lunar night time due to the saturation of the sensor from sunrise onwards. Hodges and Hoffman (1975) have clearly shown that the synodic variation of $^{40}$Ar is characteristic of a condensable gas. The slow post sunset decrease and a sharp increase during sunrise times have been shown to systematically repeat from one lunation to another. They even suggested that the sudden increase recorded even before sunrise could, in all probability, be associated with a part of the desorbed gas from the sunlit surface drifting across the terminator into the night sector. They even went to the extent of stating that, this behaviour of lateral movement of the gas itself could vindicate the $^{40}$Ar to be of lunar origin. Typically, the extrapolated/modelled daytime values were at least two orders of magnitude larger with reference to the night side measurements. Even with the limited data from LACE, Hodges etal (1973) could get evidence for a potentially varying source of $^{40}$Ar, i.e., for a varying release rate of $^{40}$Ar from the lunar interior, possibly associated with moonquakes (Hodges and Hoffman 1975). However, as stated by them, the limited data base was found to be insufficient to carry out a systematic correlation analysis with the tidal/stress related forcing. As mentioned earlier, all the above understanding is based on measurements from specific locations. There were attempts during Apollo 15 and 16 missions wherein mass spectrometers were flown in the service modules but with limited success. One of the objectives of the studies carried out by Hodges etal (1972) was to look for lunar volcanism. Though no conclusive evidence was obtained, some sample spectra obtained by them, presumably corresponding to different lunar ambience, are available in the literature wherein the $^{40}Ar / ^{36}Ar$ ratio has been shown to come down to ~3.5. No further details are available from the orbiter data even for the anti-sunward side, leave alone the sunward side.

Chandra's Altitudinal Composition Explorer (CHACE), on the Moon Impact Probe (MIP) of Chandrayaan-I, India's first mission to moon provided the first glimpse of the neutral composition of the tenuous 'sunlit' lunar atmosphere.  While CHACE provided direct evidence for water vapour and $CO_2$ in significant amount, presence of a variety of species were also reported in the literature (Sridharan etal 2010a, b).  The emphasis in the earlier work had been to report the first systematic composition measurement as the MIP descended towards the South Pole. It had also been discussed earlier that due to the virtue of the moon having a surface bounded exosphere, which implies that literally there is no interaction between the gaseous species, all the measurements from space would directly correspond to the surface processes.   This statement gets vindicated while deriving the lunar surface temperature from the altitudinal distribution of $^{36}Ar$ and its independent confirmation from spectroscopic methods reported in the literature (Sridharan etal 2010b). $^{36}Ar$ distribution is chosen to make an estimate of the surface temperature as its origin is non-lunar (in this case solar) and its distribution is expected to be uniform and global. The distribution of all other specie would be dictated by their source distribution and also the surface temperature variations.   Having independently determined the surface temperature variations, its effect on the distribution of any other species could be suitably accounted for.

As mentioned above, the nature of the surface boundary exosphere implies that every constituent has its own atmosphere covering the same volume and the variation in the relative concentration measured as the space craft raced downwards to the pole would directly correspond to the surface heterogeneity.   Once the effect of the surface temperature variations are duly accounted for, the variation in the relative concentration would then translate one to one to the variations in the source region along the ground track of the space craft. Since the complete mass spectrum is available every ~4 sec corresponding to  ~250 m altitudinal and 0.1 deg latitudinal resolution, finer aspects at higher resolution could be studied than what has been reported earlier by Sridharan etal (2010b).

In the present exercise attempts are made to study the relative variations of two of the measured species viz., $^{40}$Ar and $^{36}$Ar which are believed to have different origins and infer the processes that could be happening in the lunar interior. $^{40}$Ar is presumably produced radiogenically from the lunar interior and it showed significant spatial variation. If it is directly controlled by the processes in the lunar interior then there is a possibility of it to bear some relation with the lunar surface features like the deep craters and high elevation zones or in other words with lunar topography. Since the main orbiter Chandrayaan- I had a Lunar ranging Instrument (LLRI), the measurements have been correlated with the LLRI topography nearly in the same zone. It is believed that the combined findings give us clues on the possible spatial heterogeneity of the radiogenic activity of the lunar interior.

**Experimental details**

CHACE is built in-house centred around a standard residual gas analyzer (RGA) with an electron multiplier detector, working on the principles of quadrupole mass spectrometer and was one of the scientific experiments on the Moon Impact Probe (MIP) - a micro satellite riding piggy back on Chandrayaan-1. After attaining a stable 98 km orbit, the stand alone micro-satellite weighing ~36 kg, was separated from the main orbiter, decelerated towards an impact point on the South Pole. CHACE was switched on over $40^0$N, when the MIP was still attached to the main orbiter and provided neutral composition data corresponding to the orbital altitude of 98km till the spacecraft reached $13.3^0$ S where, after a brief manoeuvre, the MIP was released from the mother spacecraft. The neutral composition measurements were carried out all the way down to the surface of the moon that extended up to the South Pole with unprecedented altitudinal/ latitudinal resolution (~250 m/ ~ 0.1deg) along the $14^0$E meridian. From the time of switching on and till the point of impact CHACE yielded 650 spectra covering a mass range of 1- 100 amu with a mass resolution better than 0.5 amu. Each mass scan lasted ~4 sec. Nearly 50% of the spectra correspond to the orbiter altitude of 98 km and the remaining to the descent phase.

CHACE made use of a hot filament ion source and required a stabilization time of ~15 - 20 min., before absolute partial pressures could be measured; i.e., where the pre-flight inter calibration with reference pressure gauges in the laboratory could be effected. Incidentally, as revealed by the monitoring parameters (in this case the filament emission current), during the actual mission, the instrument had stabilized in ~ 15 min i.e., just before the time of its release. This fortuitous circumstance enabled partial pressure measurements right from the orbital altitude to the lunar surface. As a health monitoring exercise, five orbits before the actual release, while the space craft was hovering along $20^0$E meridian, the MIP was switched on for a short duration of ~10 min. Data from CHACE pertaining to this rehearsal phase covering $39^0$N to $9^0$N are also available. During the warm up phase of the ion source when there is a likelihood of degassing, the data may not be useful even for relative concentration studies. However, to some extent, relative concentration of noble gases, which are not due to out gassing could still be considered. Results corresponding to the descent phase of CHACE dealing with the direct detection of water vapour and also providing the first comprehensive composition measurements of gaseous species that constituted > 1% of the dominant specie, from the sunlit side of the moon, are available in the literature (Sridharan etal 2010a, b). The present paper concentrates on the relative concentration of $^{40}$Ar /$^{36}$Ar and its spatial variability along the longitudinal sector (~$14^0$E) of the MIP trajectory.

**Measurement uncertainties and out gassing issues:**

When measurements are attempted in very tenuous and unexplored atmospheres similar to the one encountered over the moon, out gassing from the probe and its surroundings become extremely important. Extensive tests have been carried out on CHACE before ascribing the measurements to the lunar ambience. In any leak proof, closed system that is continuously pumped, the ultimate vacuum level that one could reach is dictated by the out gassing from the components of the vacuum system including the measuring device and also the pumping speed/efficiency. Equilibrium would soon be attained wherein the pumping speed and the out gassing rate balance each other. The net out-gassing rate is dependent on the interior surface area, temperature and also the volatility of the substances used in the fabrication of the different components. Figure 1 depicts a sample spectrum highlighting the performance features of the mass analyzer in the laboratory. Instrument characterization has been carried out extensively at elevated temperatures (150 C) wherein both the vacuum system and also the mass analyzer were baked thoroughly for several hours under ultra high vacuum conditions. The dominant atmospheric species like $N_2$ and $O_2$ are clearly seen in the lab spectrum. The presence of $^{40}$Ar is quite conspicuous and as expected forms ~1% of the total concentration. The total absence of $^{36}$Ar and $^{38}$Ar, which are the other two stable isotopes is worth noting. Though the background spectra could be characterised with reasonable certainty, since the out gassing can never be totally eliminated, any measurement of an unknown ambience needs to be necessarily carried out in the back drop of some of these constituents. On the other hand, when measurements pertain to deep space which could be construed as an open system with infinite pumping capacity, the out gassing effects would mar the measurements only to the extent of the proximity and orientation of the inlet of the sensor to the potential source regions. The importance of the space craft out gassing on the composition measurements of tenuous atmospheres have been dealt with in detail in the literature (Schlappi etal 2010), wherein the authors have comprehensively discussed how a permanent thin gas cloud could get formed around the space craft and how the density and composition could vary with the space craft manoeuvres and payload activity. The above detailed study that pertained to one of ongoing and successful ROSETTA missions of the European Space Agency (ESA) had also identified water, organic substances from the space craft materials like insulators, lubricants and the electronics which are potential out gassing sources. It seems to substantiate the earlier inferences, whether it is the earlier lunar exploration by the Apollo series or the Mid-course Space experiment (MSX), from the near Earth space. In the latter, it had been shown that the out gassing could continue even after several years of continuous exposure to outer space. There are exceptions though, with the Cassini-Huygens probe, which also had a mass spectrometer for thermal and ionized gas (Waite etal 2004). Though the camera systems in the main probe deteriorated presumably due to space craft out gassing (Haemmerle and Gerhard 2006), the mass spectrometer experiment that had a sensitivity of ~ $10^{-12}$ torr did not register any contaminant release implying that the out gassing from the space craft, if any, had been much lower than the detection limit of the mass analyzer. This mass spectrometer also had an open ionization source similar to the one employed in CHACE of Chandrayaan-1. It should be borne in mind that all these issues including the degradation of the critical components become very important particularly in the long drawn deep space missions.

The above points are some of the prime reasons for CHACE to be opted as an experiment in the Moon Impact Probe (MIP), a micro satellite weighing just ~36kg (600x400x400 mm$^3$), as compared to ~550 kg of the main orbiter, and was designed to work only for a total duration of ~45 min. The inlet orifice of the mass spectrometer, which had a projected field of view of <1deg, was also inclined by ~90 deg to the velocity vector of the MIP. Further it should be noted that the MIP was a spinning module with CHACE mounted parallel to the spin axis. Figure 2 shows schematically the orientation of the probe in different stages as it descended towards the moon. In spite of all these precautions, because of the inherent sensitivity of the probe to measure partial pressures of the order of $10^{-13}$ torr, the out gassing from the space craft, if any, may still turn out to be significant. Since MIP is a one-shot affair lasting for a short time of ~44 min., one would not expect to notice any significant change in the out gassing rate that too when the analyzer had been exposed to deep space vacuum for more than 3 weeks and eventually, one may end up having a constant background spectra in the light of which the lunar constituents may have to be delineated. In order to have a watch on this, laboratory estimation of the out gassing from all critical components has been carried out. Water is known to be one of the toughest constituents in addition to its derivatives. The ratio of 18 amu to that of 16 and 17 amu and also, 18 amu vs 44 amu were taken as a measure of the out gassing properties of the materials. In addition, attempts were also made to get the 'finger print' masses corresponding to these materials. If these characteristic masses were detected in the actual mission, the contamination due to out gassing should be kept in mind before drawing any conclusion. The out gassing studies were carried out under ultrahigh vacuum conditions (< $10^{-8}$ torr) with the samples heated up to ~250 C. The materials characterised include the Multi Layer Insulation (MLI) with tape, the honeycomb structure out of which the MIP was fabricated, the epoxy of the honeycomb structure and the epoxies used in the electronics (Stycast 2850 FT and DC 3140). It should be borne in mind that such elevated temperatures are not to be encountered in the mission. Further, in addition to the above, one has to address the possible out gassing due to space craft operations also. Sridharan etal (2010b) have dealt with this aspect wherein they had reproduced actual spectra when i) the MIP was still connected to the main orbiter Chandrayaan-1, ii) during de-boost motor firing and iii) immediately after the release of MIP. It had been shown that the spectra had remained totally unaffected which is a clear indication for the out gassing to be a non issue in this case, some what similar to the Cassini-Huygens probe.

The total pressure as measured by the ionization gauge which is an integral part of CHACE had been consistently showing it to be in the order of $5 \times 10^{-7}$ torr, at least two orders of magnitude more than that has been anticipated based on the far side Apollo measurements. An independent estimate of the ambient pressure was also carried out using one of the instrument properties. The time taken by the ion source emission current to stabilize is strongly dependent on the ambient pressure levels. Using laboratory calibration, based on the stabilisation time taken during the mission (~900 s), the estimated ambient pressure turns out to be in the range of $8 \times 10^{-8}$ torr. Moreover, the steady state value of the filament voltage of the ionizer, which is regulated in a closed loop system in order to retain the thermionic emission current from the filament constant, is also a function of the total pressure. The stabilized value of the filament voltage of CHACE monitored during the mission, suggested the total pressure encountered to be ~ $1 \times 10^{-7}$ Torr. These two parameters (stabilization time and filament voltage) provide indirect and independent confirmation for the pressures to be significantly higher than what has been anticipated earlier. Once it is ascertained that the sunlit side pressures are indeed higher and in the range of $10^{-7}$ torr, the total absence of out gassed constituents naturally gets explained. With the base pressure of this order, with the typical saturated vapour pressure for the space craft components lying in the range of $10^{-8}$ to $10^{-13}$ torr, at the prevailing temperatures, contaminants due to out gassing could become a non issue. These have been discussed extensively by Sridharan etal (2010 b). It should be noted that the mass spectrometer while having a partial pressure detection sensitivity of better than $10^{-13}$ torr, still had a mass resolution better than 1 in 100 amu. This is considered very good for Aeronomic studies. However, it is not capable of differentiating nuclear mass defect separation of $^{40}Ar$ on one hand and $C_3H_4$ ions on the other, at 40 amu. Careful analysis of the laboratory spectra indicates no possibility of hydrocarbons being present. Though CHACE was an unsealed mass spectrometer, all the care had been taken to carry out the laboratory studies in an all metal, bakeable, ultra high vacuum system capable of reaching up to $10^{-9}$ torr. The distinctly different features of the laboratory spectra and the actual spectra from the mission stand testimony to the above arguments. In the unlikelihood sources for organics and also based on the earlier measurements during the Apollo era, the measured 40 amu, 38 amu and 36 amu are taken to represent the isotopes of Argon.

As the MIP descended from ~98km to the lunar surface it yielded the variation in the lunar atmospheric constituents with latitude/altitude which implies that at any altitude the measured concentration would be an admixture of both the altitudinal and latitudinal variation. The former is essentially dictated by the surface temperature in the case of the surface bounded exosphere of the moon and to the heterogeneity in the source regions. **Apriori knowledge of surface temperature would enable one to delineate the latitudinal variations of the atmospheric specie and this aspect will be dealt with in detail later**. The surface temperature estimates have been made independently using one of the noble gases i.e., $^{36}$Ar which is of solar origin and whose distribution is global and hence is not expected to be affected by the heterogeneity of the lunar surface. For an equatorial temperature of 400 K and considering the measured partial pressure of $^{36}$Ar at different latitudes, under 'surface boundary exospheric' conditions, the probable lunar surface temperature variation has been estimated and found to agree very well (Sridharan etal 2010 b) with the available information based on spectroscopic estimates.

The present paper emphasizes the distribution of Argon isotopes with specific emphasis to the radiogenic activity of the lunar interior. We consider only the stable isotopes of Argon viz., 36, and 40 amu for the present. Sample spectra covering the mass range of 35-41 amu alone are depicted in figure 3 to highlight the quality of data that has been obtained. The significance and variability of these isotopes could be ascertained from the figure at different stages of the mission. The first two panels of the spectra **a** and **b** correspond to times when the MIP was still attached to the main orbiter and those of **c** and **d** correspond to times when the MIP was racing towards the south pole. In the whole analysis of CHACE data, only those masses that have at least five data samples (sampled at ~ 4ms) out of the nine earmarked for every amu are considered. Having ascertained that the measured concentrations are of lunar origin, the isotopic ratios of Argon have been made use of in bringing out the difference in the radiogenic activity of the lunar interior and its possible the linkage to the lunar topography. It should be kept in mind that these are the 'first direct' measurements from the sunlit side of the moon and indicate the need for more such direct measurements.

**Results**

As mentioned above, CHACE was switched on at $40^0$ N and the measurements continued till the MIP impacted close to the South Pole; with all the information contained in 650 high resolution mass spectra. All the necessary corrections for the mass discrimination in the electron multiplier and the quadrupole analyzer have been duly applied to the raw data following the methods described by Sridharan etal (2010 b). When the ratios of the relative amplitudes are considered, stabilization of the instrument is not mandatory since within the short time span of ~250ms taken to cover the mass range of 36-40 amu the instrument characteristics are not likely to change.   Therefore even when the instrument is just getting stabilized, ratios could safely be considered.    On the other hand, when partial pressures of different species are to be dealt with, it is essential to ensure that the instrument had stabilized, which in the case of CHACE is ~15 minutes after switch on, during which time the MIP had crossed over to 20 S.  This is the reason for restricting ourselves to data beyond $20^0$ S when $^{40}$Ar alone is considered.  This point will be taken up later.  With this logic, the ratios are taken right from the first spectrum without waiting for the instrument stabilization. Results from both the rehearsal and the mission phases are depicted in figure 4, wherein 15 point running average has been shown (thick line) in addition to actual ratios in every sweep. The average ratio initially registered a value of ~ 1, over the lunar latitude of $40^0$ N, only to steeply increase to 2.25 over $12^0$ S and to remain centred around this value till the point of impact. There are large and small undulations all through, which could on occasions, reach as large as 30 % of the mean value. Interestingly the data pertaining to the rehearsal phase corresponding to $20^0$ E in figure 4 shows a value of ~1.5 at ~ $30^0$ N and it followed more or less the same trend of the mission phase. The agreement between these two phases of the mission highlights the consistency of the measurements. One important aspect of the above result is that, the ratio was consistently larger than that corresponding to the solar wind (~0.1), indicating the potential radiogenic activity in the lunar interior as had been dealt with by the earlier workers in the field and its possible variation along the ground track. Incidentally, the amu 38 which could be one more of the stable Argon isotopes also shows significantly large ratios with reference to 36 amu as compared with the solar wind values. There is no clarity especially with regard to the geochemical reactions that would result in the production of 38 amu.  In the absence of such clarity, in this paper, we are constrained to restrict the discussion only to $^{36}$Ar, and $^{40}$Ar.

In this context, some of the earlier results on the ratio of $^{40}Ar/^{36}Ar$ and their possible relation to the de-gassing history of the lunar surface under consideration become important. Wieler and Heber (2003) have exhaustively discussed about the noble gas isotopes on the moon and have highlighted how these 'parentless' gases could be used to constrain the degassing history of the moon itself. The parentless $^{40}Ar$ is used as a semi quantitative measure of the 'antiquity' of a soil i.e., the time in the past when a sample was exposed to the solar wind at the top of the regolith. Concentration of $^{40}Ar$ needs to be normalised to the length of time a sample had spent on the surface for which $^{36}Ar$ from the solar wind could be used as proxy. It is believed that the $^{40}Ar/^{36}Ar$ ratio in any sample that had been irradiated by the solar wind several billion years ago would be higher than in the sample that received the radiation later. This implies that higher the ratio of $^{40}Ar/^{36}Ar$ in any sample, the older the sample would be. A cross calibration has been provided by Eugster etal (2001) which clearly indicates that a ratio of ~13 would correspond to an antiquity of 3.7Gy while a ratio of 0.5 would correspond to 10 My. Figure 5 depicts the re-plotted cross calibration of the basic inputs given by Eugster etal (2001) and Wieler and Heber (2003). The relationship has been evolved for different samples collected from the Apollo era highlighting the spatial heterogeneity. In the present case, the $^{40}Ar$ to $^{36}Ar$ ratio after the instrument stabilization is seen to be centred ~2.25 in the southern hemisphere of the near side of the moon. This would imply an antiquity of ~1.5 Gy for the longitudinal segment corresponding to the ground track of the moon impact probe as shown in the figure. Further, going by figure 4, one notices that the ratio had remained around 1 in the northern hemisphere which indicates for the antiquity of this region to be around 600 My. An empirical relation has been worked out relating the ratio to the antiquity and highlighted in figure 5. This result opens up an interesting possibility of getting the 'antiquity' of the whole of the lunar surface by means of composition measurements from an orbiter around the moon. More systematic measurements need to be carried out in future to vindicate the above conclusions.

In the context of the above results on the antiquity of the southern hemisphere, it is interesting to compare them, at least qualitatively, with the elemental maps generated by the earlier missions. Figure 6 reproduces the Thorium and Potassium maps derived by Lawrence et al (1998) on which the ground track of the Moon Impact Probe and the trend of the ratio of $^{40}Ar$ to $^{36}Ar$ as one moved towards the pole are superposed. The purpose of depicting the MIP trajectory over these maps is to highlight the heterogeneity of the region through which the MIP passed over. Lawrence et al (1998) have produced lunar maps for thorium, potassium and iron using gamma ray spectrometry from the lunar prospector mission. These maps were generated while delineating the compositional variability and evolution of the lunar highlands which contain KREEP- rich materials (potassium, (K), Rare Earth Elements (REE) and phosphorous (P)). Since KREEP rich rocks are believed to have formed in the lunar crust-mantle boundary as the final product of the initial differentiation of the moon, their distribution would yield information on the temporal evolution of the lunar crust. Very recently, Meng Hua Shu etal (2010) have reported similar results from the Chinese CHANG'E-1 mission that carried a gamma ray spectrometer. They had confirmed the earlier results reported by Lawrence et al (1998, 2000) including the correlation between the radioactivity and lunar topology. Thus the gamma ray spectrometry could yield absolute information on large scale features though with a coarse resolution. The lunar maps generated by the earlier workers clearly demarcate the far and the near side of the moon, the former showing substantially less counts than the latter. Even in the near side, large differences exist both along the latitude and longitude. For example maximum potassium concentration has been shown to be localized and centred on $\pm 15^0$ S latitude and $0 - 40^0$ W longitude with some sort an irregular structure in the periphery (Lawrence etal 1998, Meng Hua Zhu etal 2010). It may be recalled that the ground projection of the MIP trajectory was centred around $14^0$E along the irregular periphery of the KREEP region containing a large number of small and medium craters. Since there is no mixing of the atmospheric constituents in the lunar atmosphere, one would expect the concentration at any point in the atmosphere to be directly linked to the source region below.

**With regard to the relative variation of 36 and 40 amu it should be borne in mind that both of them would show variations due to two factors 1) as the MIP descends and 2) due to the steep decrease in the surface temperature as one moves towards the pole; i.e., the measurements depict a combination of altitudinal/ latitudinal variation. As elaborated earlier, since the distribution of $^{36}$Ar is not affected by any internal source, and as such its distribution is global, it has been used to infer the lunar surface temperatures (Sridharan 2010b). While estimating the partial pressures of any other species at any location, due temperature correction has to be applied and that too only after the instrument had stabilized. Ratios however could be exempted as the instrument operating**

conditions remain the same during a typical mass scan.  For example the time difference in the present mode of operation of the instrument for sampling amu 36 and 40 is ~250 ms. And they being closely separated in the mass scale their relative variation due temperature would not be that conspicuous and variations have to be necessarily associated with their sources. The differences however, would magnify when there is an independent source for one of the species which in this case is the radiogenic activity for $^{40}$Ar.  Figure 6 depicts the ground track of MIP along with the $^{40}$Ar/$^{36}$Ar, superposed on the thorium and Potassium maps obtained by Lawrence etal (1998) to highlight the point that the projected trajectory falls on the irregular periphery of the radiogenic region. In this context it would be interesting to see whether the lunar topology has any bearing on the variations of $^{40}$Ar giving us further clues on the spatial heterogeneity of the radiogenic activity. Having known the radiogenic source of $^{40}$Ar, it is only appropriate to consider the $^{40}$Ar alone rather than the ratio.  Since CHACE yields the altitudinal/latitudinal variation of the measured specie, it would be more appropriate if the 40 Ar partial pressures are estimated right at the lunar surface before examining the correlation with the topography.   This has been a straightforward exercise knowing the fact that 40 Ar has its own scale height solely dependent on the surface temperature.   The partial pressure $P_h$ measured at any height 'h' as the MIP descended is put in the expression (1) to estimate the partial pressure over the surface.

$$P_o = P_h \exp (h\, mg/kT) \qquad (1)$$

Where, $P_o$ is the partial pressure at the surface of the moon, $P_h$ the partial pressure measured at an altitude  h, m the mass of the 40 Ar atom, g the acceleration due to gravity at the lunar surface, k the boltzman constant, and T the surface temperature corresponding to that latitude.  The $P_o$ values estimated at every instant has to be examined along with the topographical features of the lunar surface.

 In this context, data from the Lunar Laser Ranging Instrument (LLRI) on Chandrayaan-1 (Kamalakar etal 2009) become relevant.  Though LLRI, a part of the main orbiter Chandrayaan-1, was operated in a restricted mode, with more emphasis given to the Polar Regions, significant part of the moon had been covered at very high resolution during the mission's life time of one year.  The purpose of this experiment had been to glean out the lunar topography. The LLRI data corresponding to the longitude zone of the MIP ground track had been stripped out of the data base and the topological features were extracted. Since the data are of very high resolution, 60 point running mean had been found optimal to make comparison with the CHACE data. Care had been taken to restrict ourselves only to the longitude region of interest. Data from this sector had been collated from different orbits and the elevation of the lunar structures delineated and presented in figure 7. Quite a few elevated zones extending even up to 2 km and also deep craters with depths of 2.5 km have been detected by the LLRI along $14^0$E longitude. While making the comparison with the 40 Ar partial pressure the mean reference line (zero line) for the LLRI data had been kept in mind. The striking observations from figure 7 are; from ~ $42^0$ S latitude to ~$80^0$S the elevation registered a steep increasing trend from ~ -2.3 km to ~ +1.6 km and there had been a corresponding decreasing trend from ~$1.3 \times 10^{-8}$ torr to ~$3 \times 10^{-9}$ torr in the $^{40}$Ar partial pressures. Further, it **is interesting to note** that the undulations in the $^{40}$Ar are nearly **of the same scale lengths as revealed by the ranging instrument in the whole region under consideration**. This **is in spite of the fact** that the data from these two experiments were not obtained simultaneously and the LLRI derived features **were collated from different orbits** representing a longitudinal spread of ~1 deg while the instantaneous CHACE data represent a much smaller projected area on the surface of the moon. **For nearly the same elevation revealed by the LLRI in the latitude range of 20-30 S and 65-75 S the partial pressure varied from ~~$1.3 \times 10^{-8}$ torr to ~$5 \times 10^{-9}$ torr and this itself could be considered as indicative of spatial heterogeneity. The results indicate a plausible link between the surface features and the atmospheric parameters and this warrants more detailed investigation. Though the presented results are from a one shot experiment and need to be confirmed by future concerted studies,** CHACE results that complement results from other experiments like gamma ray spectrometry have brought out finer aspects thus opening up interesting possibility for future missions.

**Conclusion**

The in-situ measurements of the ratio [$^{40}$Ar / $^{36}$Ar] and the laser based topography data bring out the heterogeneity in the radiogenic activity of the lunar interior, with a very high spatial resolution in addition to providing semi quantitative antiquity of the region falling under the ground track. This opens up the interesting possibility of using the ratio of the noble gas isotopes to establish the antiquity of the whole of the moon by suitably planning the lunar atmospheric composition measurements in future missions. An instrument like CHACE with a larger mass scale extending up to at least 300 amu in an orbiter would enable mapping of the radiogenic activity at very high resolutions. Further, the significantly larger concentration of the Argon isotopes indicates the need for a revisit of the estimates of the source potential and also the loss processes. The proposed follow up mission from India, Chandrayaan-II would carry an instrument of the kind (Chandra's Atmospheric Composition Explorer-2 : CHACE-2) with an extended mass range in the orbiter, and hopefully would yield high spatial/temporal resolution data that would enable us to refine our understanding of the lunar interior in general and the lunar atmosphere in particular.

**Acknowledgements**

The authors gratefully acknowledge all the help by the MIP-Chandrayaan team and the various technical groups in VSSC in the execution of the project. The support from the technical staff of SPL is duly acknowledged. Useful discussion with Prof. J. N. Goswami, principal scientist of Chandrayaan-1 and SVS Murthy of PLANEX program is duly acknowledged. RS duly acknowledges the CSIR for the CSIR emeritus position and the director PRL for hosting the position. This work is supported by the Dept. of Space, Govt. of India.

**List of figures**

Figure1: A typical laboratory spectra of the CHACE instrument under ultra high vacuum conditions after a bake-out at elevated temperatures of 150 deg C high-lighting the background constituents of instrumental origin at a pressure level of $10^{-8}$ torr similar to what has been encountered in the lunar ambience.

Figure 2: Schematic representation of the separation of the moon Impact Prove (MIP) from the main orbiter Chandrayaan-1. The separation is effected after an initial manoeuvre of the main craft. The inlet orifice of CHACE is mounted along the spin axis of the MIP which is ~90 deg with respect to the velocity vector shown by an arrow.

Figure 3: Sample spectra covering the mass range of 35-41 amu showing clear cut presence of $^{40}$Ar and $^{36}$Ar. Coherent addition of 20 consecutive spectra and averaging have been made and the representative latitudes that cover both the north and southern hemispheres have also been marked.

Figure 4: Ratio of $^{40}$Ar /$^{36}$Ar as obtained by CHACE during MIP rehearsal (along $20^0$E meridian), and actual mission (along $14^0$E meridian) along with $^{38}$Ar /$^{36}$Ar . 15 point moving average is resorted to in order to show the structures. The raw values of the ratio acquired during the actual mission reveal very high values at certain locations.

Figure 5: Redrawn 'antiquity' estimates of the lunar samples along with their corresponding $^{40}$Ar /$^{36}$Ar ratios using the basic data from (Eugster etal 2001) along with the present estimates for both the northern and southern hemispheres.

**Figure 6: Ground track of MIP superposed on the Potassium and Thorium maps generated using the LRO data (Lawrence et al 1998). The ratio of 40 Ar /36 Ar is also shown along with to give a feel of the heterogeneity that could be expected.**

Figure 7: $^{40}$Ar partial pressure trend co-plotted with the elevation data from the Lunar Laser Ranging Instrument (LLRI), in the main orbiter Chandrayaan-1 corresponding to the MIP ground track longitude of $14^0$E. Attention is drawn to the anti-correlation trend and also the similarity in the scale lengths (pl. refer text for details).

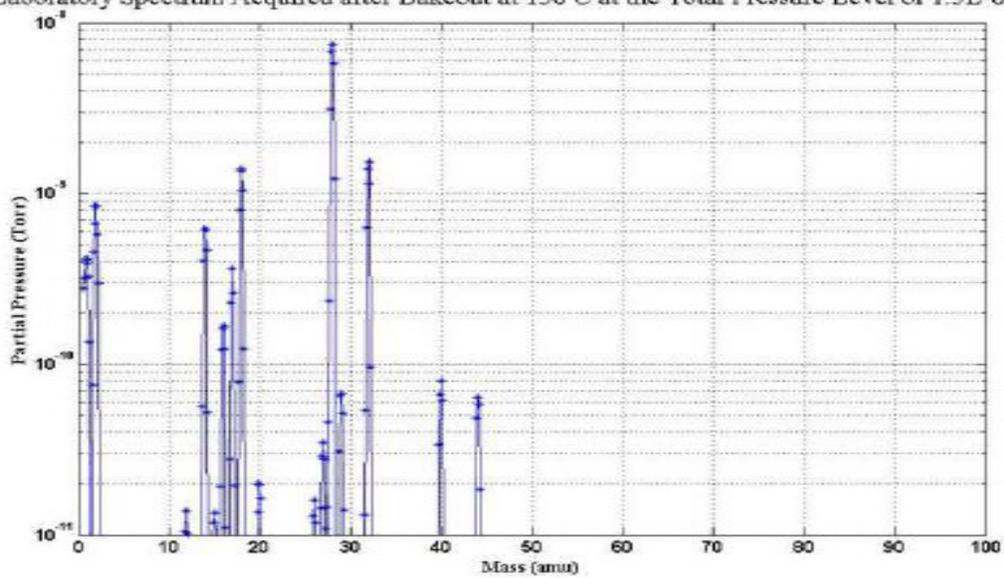

**Figure 1**

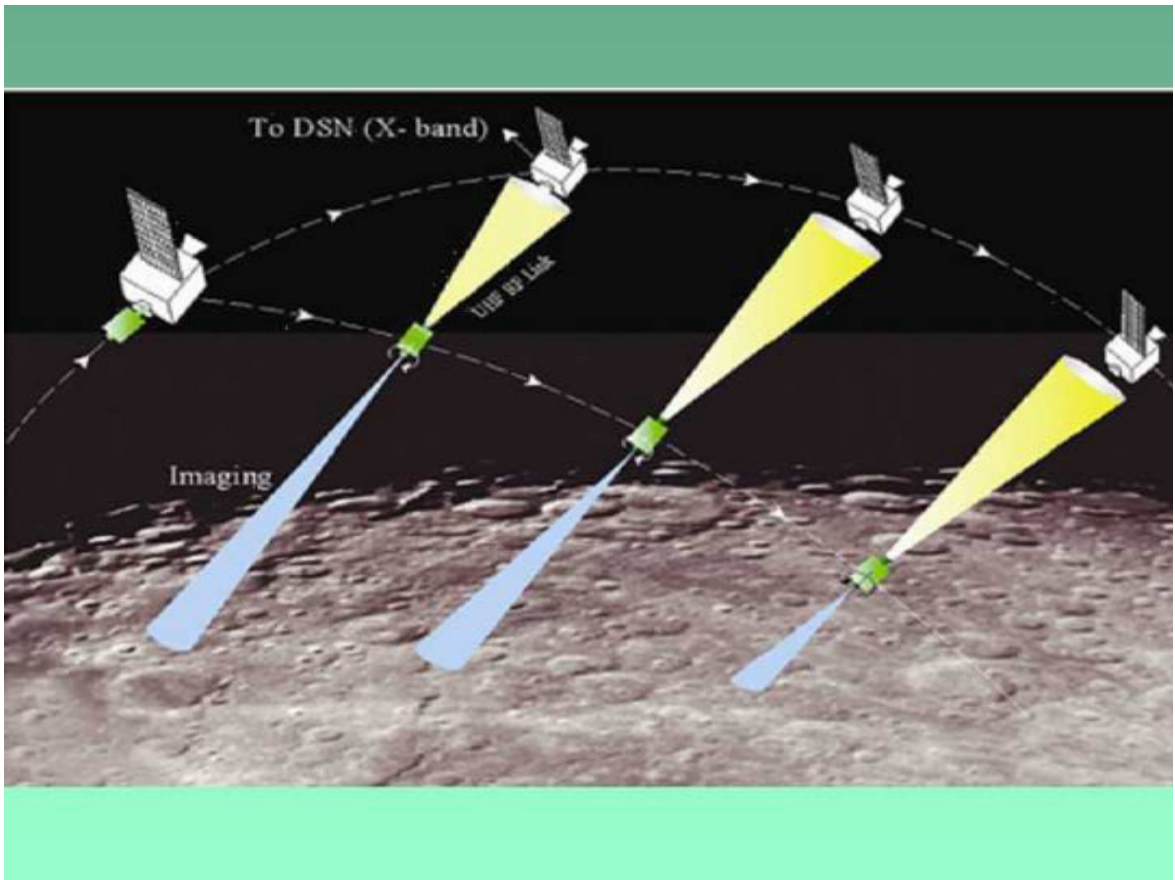

**Figure 2**

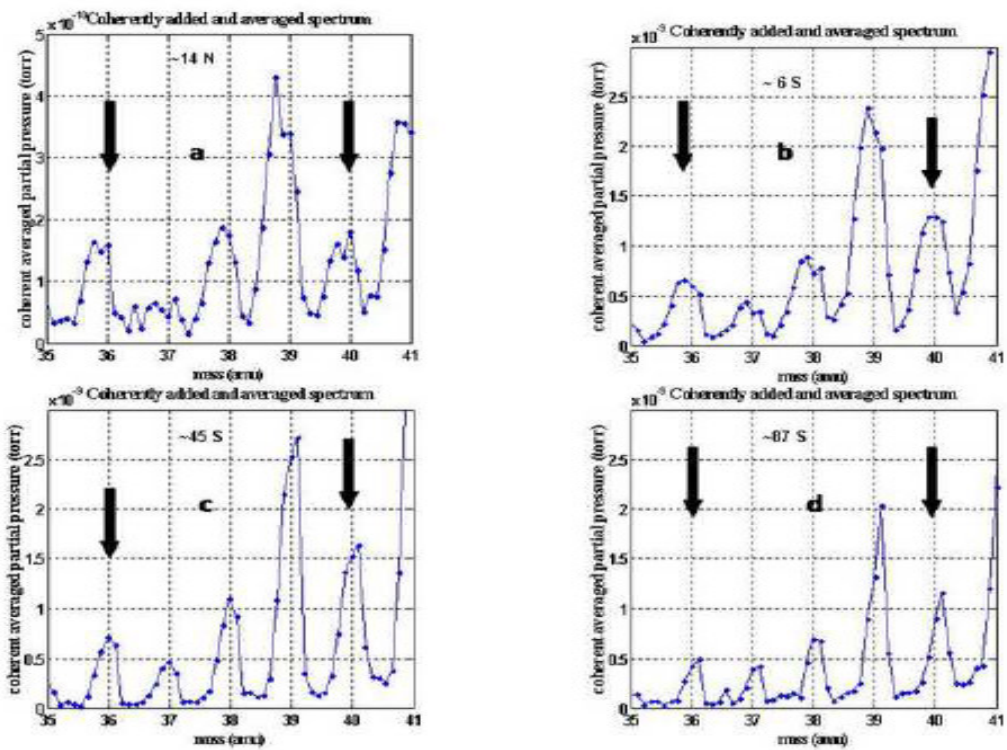

**Figure 3**

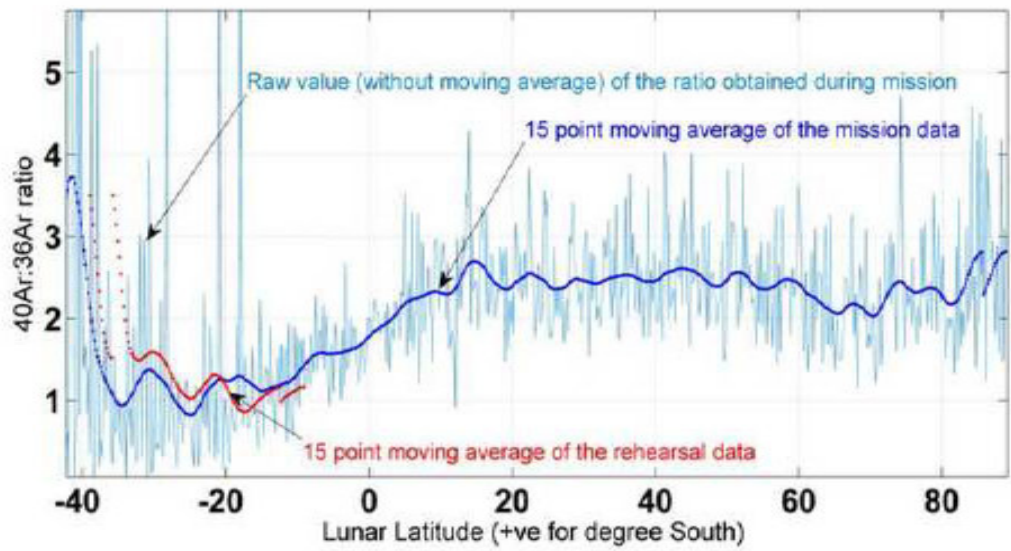

**Figure 4**

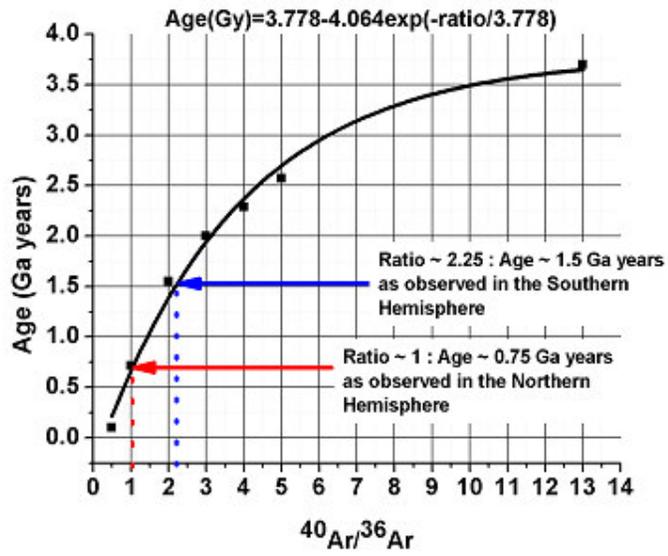

**Figure 5**

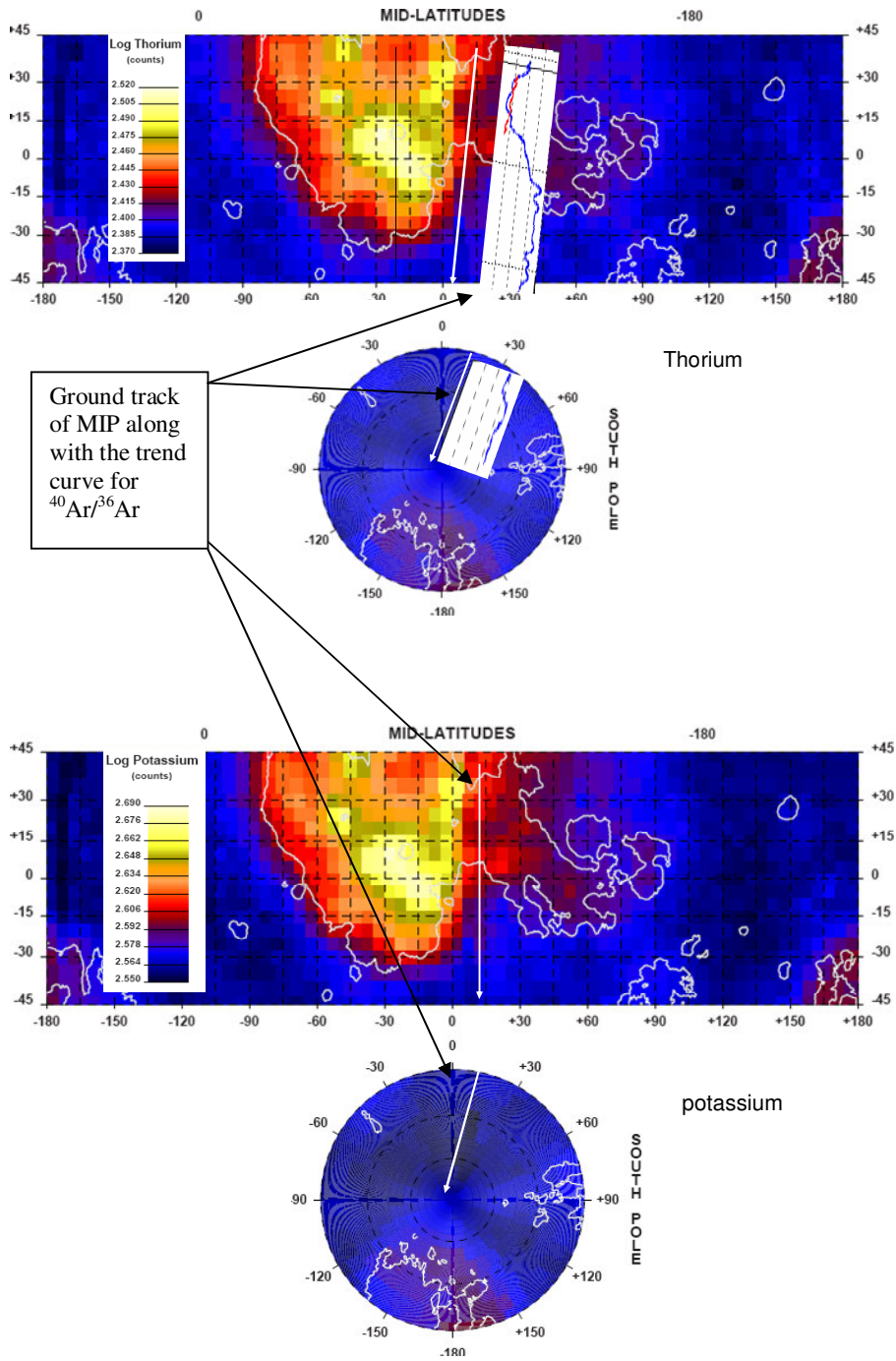

**Figure 6**

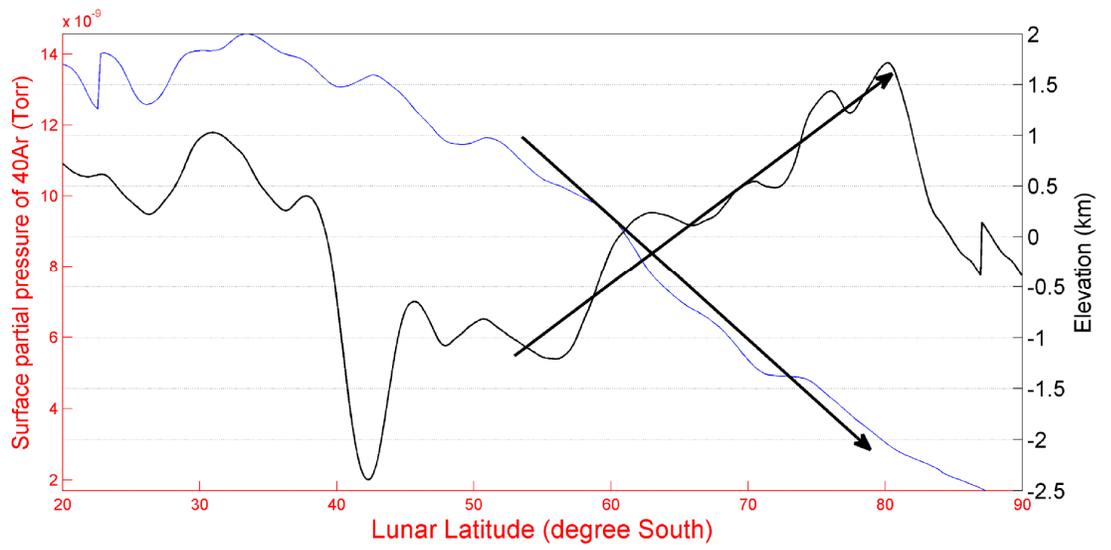

**Figure 7**